# Analysis of Research in Healthcare Data Analytics

## Mohammad Ahmad Alkhatib

School of Systems, Management and Leadership
University of Technology Sydney
Sydney, Australia
Email: Mohammad.A.Alkhatib@student.uts.edu.au

## Amir Talaei-Khoei

School of Systems, Management and Leadership

University of Technology Sydney
Sydney, Australia
Email: Amir.talaei@uts.edu.au

## Amir Hossein Ghapanchi

School of Information and Communication Technology

Griffith University

Queensland, Australia

Email: a.ghapanchi@griffith.edu.au

## Abstract

The main aim of this paper is to provide a deep analysis on the research field of healthcare data analytics., as well as highlighting some of guidelines and gaps in previous studies. This study has focused on searching relevant papers about healthcare analytics by searching in seven popular databases such as google scholar and springer using specific keywords, in order to understand the healthcare topic and conduct our literature review. The paper has listed some data analytics tools and techniques that have been used to improve healthcare performance in many areas such as: medical operations, reports, decision making, and prediction and prevention system. Moreover, the systematic review has showed an interesting demographic of fields of publication, research approaches, as well as outlined some of the possible reasons and issues associated with healthcare data analytics, based on geographical distribution theme.

## 1   INTRODUCTION

Today's healthcare industries are moving from volume-based business into value-based business, which requires an overwork from doctors and nurses to be more productive and efficient. This will improve healthcare practice, changing individual life style and driving them into longer life, prevent diseases, illnesses and infections.

Over the last few years, healthcare data has become more complex for the reason that large amount of data are being available lately, along with the rapid change of technologies and mobile applications and new diseases have discovered. Therefore, healthcare sectors have believed that healthcare data analytics tools are really important subject in order to manage a large amount of complex data, which can lead to improve healthcare industries and help medical practice to reach a high level of efficiency and work flow accuracy, if these data analytics tools applied correctly, but the questions are how healthcare organizations are applying these tools today, and how to think about it's future use? Also, what are the challenges they face when using such tools? And finally, what are the innovations can healthcare add to meet these challenges?

This paper aims to proof that healthcare data analytics techniques are not efficient enough and suitable anymore these days in order to manage big data issue and improve healthcare data analytics due to the rapid growth and evolution of technology. Moreover, it's also aims to promise professionals of a better quality of medical results, as well as reduce time needed to analyze healthcare data by keeping systems up to-date and sorting medical data in a logical structure along with accessing and retrieving patient's historical data fast and smoothly. Stakeholder 2 (Doctors and nurses.

In order to meet our goals, the proposed study is going to discuss critically weaknesses, disadvantages, problems and gaps of traditional healthcare data analytics techniques in order to manage healthcare big data. Also, it's going to develop a healthcare data analytic technique that will promise for a better medical practice and healthcare data predictive analytics based on filling gaps of traditional healthcare data analytics techniques and overcoming its problems.

As study in the area of healthcare data analytics, hospitals and clinics are looking for a new data mining techniques that will suite evolution of information technology and analyze a huge amount of complex data. The proposed technique is recommended rather than offered, since it will facilitate and enhance healthcare practice, by enabling systems to use data and analyze it efficiently and smoothly, because it will fill the gaps of previous techniques used in the hospitals, handle big data issue and avoid data loss, which will lead to improve care, assist diseases prediction and prevention systems and reduce cost. This technique is promising a better results and more benefits if it's applied correctly and properly.

Professionals (doctors and medical stuff) will be benefited for sure and they will use proposed technique, since it will reduce their time and efforts, therefore this technique focused also on adding a true assistance to their job to run smoothly as its really stressful and valuable, so they need a technique that facilitates their job and save their efforts such as: retrieving historical and old data quickly, sorting data in a logical structure way and keep it up to date, which will help them to discover hidden patterns and extract information effortlessly and efficiently. Moreover, professionals will be satisfied as they will touch that this technique will provide and additional source of knowledge to make a better decision (support decision making process) which is really needed to improve individuals' healthcare and increase their reputation.

## 2　LITERATURE REVIEW

### 2.1 Information System & Information Technology in Healthcare Sectors

The healthcare sector is widely considered as one of the most important industries in information technology (Wager 2005). More and more, information technology has been considered as a practice that facilitates healthcare performance through using data and information efficiently within the healthcare sectors. Therefore, Wager et al (2005) said that in order to understand the relation between information technologies and healthcare, we first need to understand what are the technologies used in healthcare.

Information technology functions have developed over the last few years not only as a technology services provider, but also as a strategic provider that develops and integrates industries' infrastructures to facilitate and ensure quality of service (LeRouge et al 2007).

In the mid-80,'s information technology changed the healthcare industry and brought many benefits when they used microcomputers, which were a small in shape, fast and very powerful for that time. Moreover, this allowed hospitals to develop clinical applications for various medical care settings. As a result, hospitals started to purchase and adopt information systems in the healthcare industries, and after that, challenges began to emerge when professionals tried to integrate data among these systems (Wager et al 2005).

However, Bhattacherjee and Hikmet (2007) and Castro (2007), granted that information technology has improved healthcare industries, but they also highlighted some of the difficulties related to the use of information technology in healthcare sectors, as they noticed that it is hard to implement information technology in small clinics and organizations, with high costs due to reduced efficiencies of scale. Therefore, IT implementation requires long term training and retention of skilled professionals.

On the other side of the debate, (Abbott and Coenen 2008) believe that information systems and information technology occupy a high position in improving healthcare industries in general, and in electronic healthcare record (HER) in specific; for the reason that implementing such technologies can save costs and times associated with daily hospital data records, such as patients schedules and billing. This is in addition to improving healthcare performance and efficiency by eliminating manual data records and paper work, and alongside smooth and flexible tracking of patient details.

## 2.1 Healthcare Analytics & Data Mining

Data Mining is described as a process by which data is gathered, analysed and stored in order to produce useful and high quality information and knowledge. This term also includes the way of how this data is gathered, filtering and preparation of the data for use and finally the processing of data to support data analytics and predictive modelling (Russom 2011).

### 2.1.1 Data collection

The first stage of data mining is the process of gathering and collecting data. However, even before gathering the data, ideas and plans should be assumed to decide which data should be gathered in order to collect specific data as desired and use it efficiently (Lamont, 2010).

Furthermore, Chordas (2001) added that a lot of projects fail and exceed estimated costs because of poor quality of gathered data which can result from poor data cleaning.

## 2.2 Healthcare Sectors & Big Data Analytics

### 2.2.1 Big data storage and management

One of the most important elements in dealing with and managing data is to know where and how this data will be stored once when it is collected. The traditional methods of storing and retrieving such data are not efficient anymore, since it was structured and stored in data warehouses and relational databases, after extracting and loading it from different outside sources. However, this data is transformed and classified before being ready to use and function (Bakshi 2012).

Furthermore, Herodotou et al (2011) agreed with Bakshi (2012) when he said that there are many numbers of data sources now and that a huge amount of data has become available, so this growth of data will absolutely require an agile database which can deal with the data logically and through data synchronization in order to adapt to the rapid data evolution.

On the other hand, Plattner and Zeier (2011) stated that databases only manage server memory data, therefore eliminating the option of managing other storage devices such as: disk and compact drivers. Accordingly, this will reduce the efficiency of database performance and real time response during the time.

### 2.2.2 Patients Role in Healthcare Analytics

This section is concerned about how individuals (and patients in specific) can improve healthcare analytics through understanding the small and personal data, as well as educate themselves in how to collaborate with the healthcare data analytics to reach a high level of efficiency and accuracy (Luciano 2013).

Swan (2012) was discussing the same point when he identified the term "citizen science", where non-professional and educated individuals are skilled enough to conduct and support healthcare analytics system. Accordingly, this will require organizations to train individuals how to follow up and track their health information, as well as self-monitoring.

Principally, to perform good data analytics, first of all we should teach individuals how to understand and realize the importance of dealing with such data, for instance how to deal with breast cancer (Hanoch 2012). However, Miron et al (2011) believed that whatever and how much our patients are educated and skilled to provide us with the data we expect, medical professionals still highly need to test and clarify this data to consider it and keep it on record. Also, he added that once when the data has been tested and clarified, we then need to find out how to change an individual's behavior, starting with parents and guardians who are responsible for raising their children.

However, Kim (2013) says that sometimes being motivated for change and in understanding of information are not enough. Furthermore, patients should identify the risks and detect where to change, for instance; some patients know that they have a high level of blood pressure but they don't know how to deal with it and control it: should they change drugs? Change eating habits and life styles? Do more exercises?

Turner (2011) agreed that social media and internet applications have a big influence on collecting patient information through filling and completing some online forms in order to keep track of their state of health, as well as to provide the suitable advice and treatment when needed.

Moreover, patients can share some information with other patients, so they increase their knowledge, background and awareness in the healthcare analytics sectors regarding their conditions. Finally, patients who share their symptoms, diagnoses and results with others can gain benefit from the ability to understand their health conditions by comparing them with other patients (Brownstein & Wicks 2010).

### 2.2.3 Connectivity between Healthcare Analytics System and Individuals (Medical Staff and Patients)

Connectivity approaches generate thoughts and ideas from connected networks of minds and leverages prior experiences with the utilization of technologies in our everyday life. (Siemens 2004)

Moreover, McHorney (2009) has added that healthcare analytics is not solely regarding technology and the knowledge however; it is also regards how much individuals are attached to and familiar with medical care systems and their personal skills such as ability to learn and adopt such systems in their life, as different people have different attitudes and reasons for not accepting such technologies, especially older people.

### 2.2.4 Healthcare Predictions and Decision Support System (DSS)

Healthcare prediction is another data analytics method focusing on reducing future medical costs. Predictive technique uses patient medical history to evaluate all the potential health risks and predict a future medical treatment in advance (LexisNexis 2015).

(Loginov et al 2012) stated that by retrieving and reviewing past patient details, information and diagnoses from the databases, predictive methods can take a place through forecasting, reducing time and costs.

Parkland hospital in Dallas, Texas has launched a predictive system which scans all patients' details and information to identify potentials risks and outcomes. As a result, the hospital has saved more than half a million dollars, especially in heart failure and disease predictions in terms of performing patients' monitoring and avoiding future complications (Jacob 2012).

### 2.2.5 Role of Predictive Analytics in Medical Healthcare

Predictive analytics supports healthcare sectors to achieve a high level of effective overall care and preventive care, as predictive systems' results allow treatments and actions to be taken when all the risks are recognized in early stages, which aids for minimizing costs. (Conley et al 2008).

Furthermore, Obenshain (2004) said that patients can also work and support medical care by following up and updating their medical status, so they can get the necessary treatment at the right time.

The technology era has added significant value to the healthcare decision support system, since decision making systems in healthcare care sectors can be enhanced by focusing on patient diagnoses, behavior, and prevention in order to reach a high level of care and improve healthcare economics (Cannon & Tanner 2007).

### 2.2.6  Healthcare Prediction Examples

In the healthcare sectors, predictive analytics can be achieved in many ways such as; a medical care delivery success, which can be achieved by using a model that proposes algorithms in order to assist medical treatment for interacting diseases, which can reflect in capturing patient's behavior and interactions. Another method of using predictive analytics regards how to use applications and software services alongside the electronic clinical records to analyze diagnosis and confirm outcomes in order to provide the correct treatment for the right patient at the right time (Lamont 2010). Moreover, Imamura has found that the association diagnostic approach can effect efficiently in extracting desirable information from huge databases (Imamura et al 2007).

### 2.2.7  Financial Factors in Healthcare Predictive Analytics

The most significant and obvious result of using such technology within the healthcare sectors is its results on costs. Because of cost, information is one of the main aspects that have a big effect on the cost of healthcare predictive analytics. Medical care systems have focused on increasing healthcare analytics performance as well as minimizing the cost by simplify unstructured clinical record and reducing irregular information. Consequently, large quantities of information then will be managed and controlled smoothly and efficiently (Bertsimas et al 2008).

Predictive analytics can assist to avoid and reduce inaccurate prediction costs plus time for the reason that it makes the data sourcing cost lower by specifying the desired and necessary data only, since the data is simplified, standardized and exists in historical clinical databases (Bradley & Kaplan 2010)

### 2.2.8  Healthcare Analytics & Real Time

Murphy (2013) believed that real time analytics produces more accurate results and information, since it evaluates current patients' history and conditions, therefore investigating patients' diagnosis correctly and offering the best treatment.

Real time monitoring techniques guarantee to keep data up to date and increase the quality of information, as assumed so by Taylor (2010). She believed that real time matters in healthcare analytics are very significant for the reason that it generates accurate results, such as where diabetes patients can recover if their ailment is discovered and treated correctly in the earlier stages. Walker et al (2012) agreed with that however they also highlighted some of its disadvantages, such as high cost, its high required level of training and long time to complete.

## 3  METHOD

The objective of this paper was to conduct a review, which encourages professionals, doctors, medical staff and patients to adopt and utilize technologies in order to assist healthcare analytics and improve decision making process in our everyday life.

Our method has followed three steps: 1) searching for initial and related studies, 2) Relevance appraisal and evaluation, and finally extracting data. The next sections will explain these steps briefly.

Searching for initial and related studies: the first step in order to find the articles was to specify and identify main keywords (Dieste et al. 2009). A survey was conducted to study relevant papers published since 2010 in the information system field in general and healthcare analytics and medical decision support system in specific. This study has found that most relevant keywords to "healthcare analytics" and "data mining" used with technology to support medical information systems.

The following searching phrases were used and structured in searching for relevant papers in many different databases – i.e. the relevant and related papers should contain in its titles, keywords, abstract or full text the word "healthcare" along with any of "analytics", "metrics", "data mining", "big data" or "decision making" see Table 1.

| Group A | Group B |
| --- | --- |
| **E-Healthcare** | Analytics |
| **Medical Practice** | Metrics |
| **Health** | Decision Making |
| **Clinical** | Prediction |
| **Hospitals** | Big Data |
| **Care systems** | Data Mining |
| **Wellness programs** | Business Intelligence |

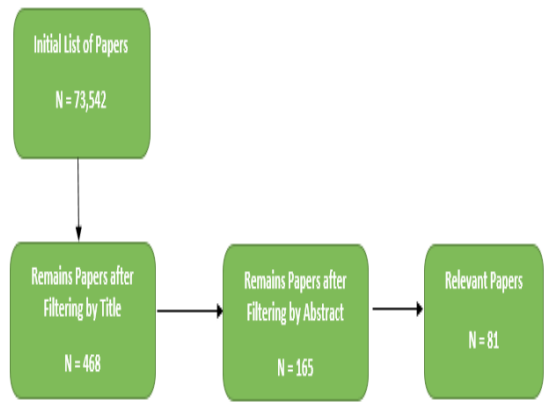

*Table 1: Main Keyword*                                *Figure 1: Search Process*

As a result, searching techniques relied on choosing any key word from the (group A) **and** linked it with any word of (group B) to form searching keywords statement, such as ("E-Healthcare" **OR** "Medical Practice" **OR** "Health" **OR** "Clinical" **OR** "Hospitals" **OR** "Care Systems" **OR** "Wellness Programs") **AND** "Analytics" **OR** "Metrics" **OR** "Decision Making" **OR** "Prediction" **OR** "Big Data" **OR** "Data Mining" **OR** "Business Intelligence".

Once the keywords were identified, 7 online databases were searched to find the initial list of the studies. In the search, titles, keywords, abstract and full text were considered and the search was limited to studies published since 2010, inclusive.

The databases were searched over multiple subjects and returned total of 73,542 articles (see Figure 1). This study found some of the papers are indexed by multiple databases. As it shown in Table 2, total number of the papers after deducting the repeated papers was 81.

| Name of Database | Initial list of papers | Papers Filtered By Title | | Papers Filtered By Abstract | | Papers Filtered By Text | |
|---|---|---|---|---|---|---|---|
| | | Number of found articles | Number of repeated articles | Number of found articles | Number of repeated articles | Number of found articles | Number of repeated articles |
| Google Scholar | 37,000 | 72 | 11 | 47 | 11 | 25 | 9 |
| IEEExplor | 6,791 | 103 | 5 | 39 | 5 | 20 | 5 |
| ACM Digital Library | 2041 | 49 | 6 | 21 | 4 | 7 | 2 |
| ProQuest | 16,139 | 133 | 8 | 44 | 5 | 31 | 4 |
| Scopus (Elsevier) | 1,380 | 104 | 3 | 29 | 3 | 14 | 3 |
| Springer | 8568 | 43 | 9 | 15 | 4 | 6 | 1 |
| Wiley InterScience | 1623 | 11 | 5 | 7 | 5 | 3 | 1 |
| **Total** | **73,542** | **468** | | **165** | | **81** | |

*Table 2: Initial list of studies*

Relevance and related papers step was completed by filtering relevant papers from the initial papers list and eliminating undesired ones, by filtering papers first based on titles, abstracts and finally full texts. Essentially, eliminated papers and articles have been done following exclusion criteria according to many main factors such as:

- Did not focus on utilizing technologies in order to improve healthcare analytics.
- Did not provide any applicable methods and experimental evidence.
- Were in different languages not English.
- Were not relevant enough and kind of old, which cannot be applied in these days.
- Were not available anymore.

Finally, extracting and analysis of data step comes after specifying and identifying related papers, so from the obtained papers now it can be extract 1) the year of publications, 2) methods and tools have been used professionals, 3) investigate if these tools and methods are still useful now a days or not, 4) what kind of problems that have been solved using these tools, 5) role of patients' in order to assist healthcare analytics and improve medical care decision process, 6) application areas, 7) research approaches and 8) data availability and geographical area of data gathering.

# 4   RESULTS

## 4.1   Fields of Publication

This systematic review has found Information Systems with 43 papers and Healthcare with 31 papers as most active communities in the research related to the healthcare analytics topic, however 7 papers were published related to the healthcare analytics in Computer Science. Figure 3 shows that authors focused on information systems and healthcare fields more than computer Science.

As it shown in Figure 2, academic papers related to the healthcare analytics and decision making were mostly published in information systems and healthcare for the reason that most studies recently have focused on improving healthcare analytics using data mining and business intelligence techniques, however a few were published in the field of computer science. This could be because of computer scientist were dealing with old traditional methods and trying to solve general issues using these methods rather than suggesting new techniques due to the evolution of technology in these days, as well as most of articles were trying to use old methods and involving only doctors and professionals in healthcare analytics process in order to improve clinical and hospitals performance as an interest topic in computer science, paying no attention for the importance to involve patients in that.

## 4.2   Application Areas

This paper found that healthcare analytics papers has focused on six main areas: (a) healthcare decision making (b) predictions of diseases & patient sickness preventions (c) clinical delivery (d) clinical operations, performance monitoring & reporting (e) Improved diagnoses, treatment and results and finally (f) Healthcare information exchange. However, there was some studies focused on financial and supply chain management, but due to the small number of publications in these areas, we removed them from this Figure.

As it shown in Figure 3 most attention was paid to improve healthcare analytics performance and results, therefore the Figure below illustrates that most publications focused healthcare prediction and preventions, decision making process and healthcare treatment and monitoring.

For many researchers, the main factors of reaching high level of healthcare analytics were in simplifying unstructured clinical records, as well as capturing patient's behavior and encourage individuals to educate themselves, as well as keep following ups, by adopting technologies and internet services and applications, for example social networking sites & social media which may allow them to keep updated and connected with other patients worldwide, sharing their health information and supporting each other in order to improve healthcare analytics and reduce costs.

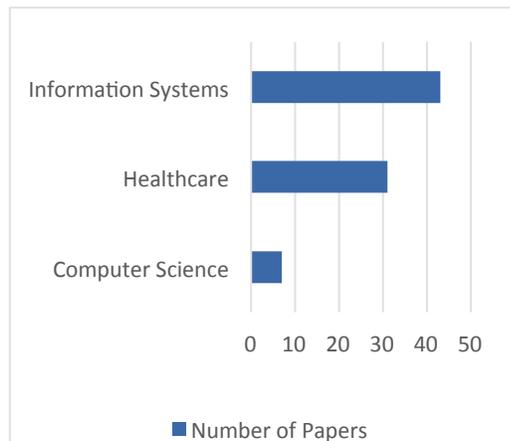
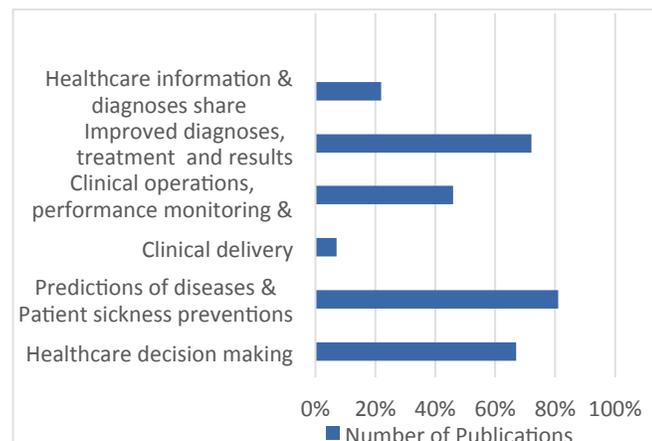

*Figure 2: Fields of Publications*　　　　　*Figure 3: Application Areas*

### 4.3 Healthcare Data Analytics Platforms and Tools:

Comparing between traditional analytics and advanced analytics, traditional analytics is focusing on business intelligence, operational research and data mining. However, advanced analytics is focusing on descriptive, predictive and optimization (Raden 2010).

The paper has come up with different tools and techniques that would improve healthcare data analytics in order to support descriptive, predictive and prescriptive healthcare data analytics the first tool is:

- **Advanced Data Visualization (ADV):** ADV is different from other standards bars and line chart, since it can scale its visualization for millions of data points, also can handle different data types. ADV is easy to use and supports analysts to explore data widely. ADV can reduce quality problems which can occur when retrieving medical data for extra analysis. Moreover, ADV can offer rich results and fluid interactions in order to reveal clinical hidden patterns in the data. (Powell 2014; Wongsuphasawat et al. 2011)
- **Presto:** is a distributed SQL query engine used to analyze huge amount of data that collected every single day. There is nothing better for healthcare sectors to find such a product which can handle a large amount of data that will come into the system. Data can take many hours and even days to be analyzed, but with Presto data now can be analyzed in just seconds or minutes. (Wulff 2013)
- **Hive:** is one of the programs developed in order to handle large amount of data, it's is not processing and analyzing data quickly as presto, however Hive does all excel tasks efficiently that don't need for real time performance, due to this companies can use both Presto and Hive for best performance, since presto can access data stored on Hive. (Capriolo et al. 2012)
- **Vertica:** program is very similar to Presto, but less expensive for the reason that Vertica eliminated costly architecture that used to associate with large amount of data. Also, Vertica has the feature of scalability which means it can cover hospital's data and analytics no matter how that data is big. Vertica can improve healthcare by reducing operational costs, accelerating medical reports and analyzing patients' patterns. (Vertica 2010; HP 2013)
- **Key Performance Indicators (KPI):** is a strategy evaluates in how company is executing its strategic vision. KPI can improve quality of medical healthcare for patients who are susceptible to hospital conditions when KPI used to specify significant indicators to be monitored and corrected, as well as identifying weaknesses. Also, KPI can use electronic medical record data to identify human practice and interventions. (Al-Azzawe 2014)

- **Online Analytics Processing (OLAP):** can improve healthcare system by performing statistical calculation very fast through hierarchal and multidimensional organized data, and can increase data integrity checking, quality control and reporting services. OLAP has the ability to improve healthcare decision making system by giving a better tracking of medical records and diagnoses. (Peši et al. 2009)
- **Online Transaction Processing (OLTP):** is similar to OLAP, but it is designated to process patient care operations, such as patient registration, hospital documents and results review. (Ledbetter and Morgan 2001)
- **The Hadoop Distributed File System (HDFS):** HDFS enhances healthcare data analytics system by dividing large amount of data into smaller one and distributed it across the other systems. Eliminating data redundancy, since HDFS has such feature built into storage layer which makes professional to focus on other responsibilities. HDFS can add a value through helping medical purposes in order to personalized treatment planning, assisting diagnosis, monitoring patient's signs and fraud detections. (Shvachko et al. 2010; Datastax 2013; Nori 2014)
- **Casandra File System (CFS):** CFS is also distributed system like HDFS, however CFS is a designated system to perform analytic operation with no single point of failure. (Datastax 2013; Lakshman and Malik 2010)
- **MapReducing System:** MapReducing system breaks Task into subtasks and gathering its outputs, as well as it enables many of the most common of operational calculations to be performed efficiently in a large amount of data. MapReducing system keep tracking on each server when tasks is being performed. The key strength of using MapReducing is the high level of parallelism, since many tasks can have performed at the same time if it's not waiting for other tasks results. (Dean and Ghemawat 2008),
- **Complex Event Processing (CEP):** CEP has come recently to the healthcare sectors, which means an event of changing in state, for instance suppose a patient gained more weight and moved from obesity to morbidity obesity. Now complex patient event processing will detect this new pattern and add it to the patient's events and relate it with being diabetic, which means that complex event processing is relating and linking events to the real time, as well as that will enhance EMR and HER systems. (Webster 2011 )
- **Text Mining:** Text Mining tools can be used and add a value in healthcare in terms of analyzing clinical records from the hospital emergency departments of physician response on call, as a similar complaints called the emergency department and were treated differently depending on the person who answered the phone. Such matter can effect in the quality of healthcare, as well as costs. Therefore, text mining can offer a treatment plan which will develop some standards and protocols to understand this matter. (Raja et al. 2014)
- **Cloud Computing:** Cloud computing has increased hospital flexibility in order to respond for dynamic changes and latest medical updates, in addition to demonstrate a great healthcare value by reducing costs, increasing productivity and security and improve data analysis with minimal management effort or service provider interaction. Cloud computing reduce strain which caused by huge amount of clinical data. One of the cloud innovations is Phillips Healthsuite platform that manages healthcare data and support doctors and patients. Phillips Healthsuite platform stores a huge amount of clinical and patient data which can be used directly in the future as an actionable data, a source of diagnosis analysis and disease prediction and prevention to increase patient care. (IBM 2011; Philips 2015)
- **Mahout:** Mahouts is an apache project aims to generate applications that supports healthcare data analytics on Hadoop systems. (Hortonworks 2015)
- **JAQL:** JAQL is a functional query language aims to process large sets of data. JAQL facilitates parallel processing by converting high level queries into low level ones. JAQL assists and works well with MapReducing. (Beyer et al. 2011)
- **AVRO:** AVRO facilitates data encoding and serialization, which improves data structure by specifying data types, meaning and scheme. (Good 2013; Confluent 2015)

### 4.4 Year of Publication

This section presents the distribution of studies and statistical trend by the year of publication. Figure 4 shows clearly that publications on healthcare analytics related topics started in the year of 2010 with only 2 studies (2% of publications). However, 2011 there was a slight increase in the number of studies by 7 studies with 7% of publications, followed by a massive increase in 2012 of 33 studies or 40% of publications, on the other hand in 2013 we noticed a big decrease by number of publications, since 20 studies had published in 2013 (25% of publications), moreover 2014 and 2015 publications also were decreased by (8 studies 9%, 11 studies 13%) respectively. we believe that this decreasing happened for the reason that healthcare data analytics is very expensive and need a lot of money to implement its projects, as it shown in the geographical distribution in Figure 7, most papers in healthcare data analytics has published in the US, and since healthcare analytics costs is high, and regarding the US poverty, as Gabe (2015) stated that US had witnessed a change in the poverty, as Washington state poverty increased from 13.5% – 14.1%, also New Jersey and New Mexico, we think this could be a good reason to kill researchers motivation and keep them away from healthcare data analytics. Moreover, healthcare analytics requires long time to adopt and not highly secured, as well as these systems have medical history data and should keep it highly secured and confidential. So, in my opinion we would assume that these kind of issues would affect negatively in researchers' spirit to go in more and more in healthcare analytics.

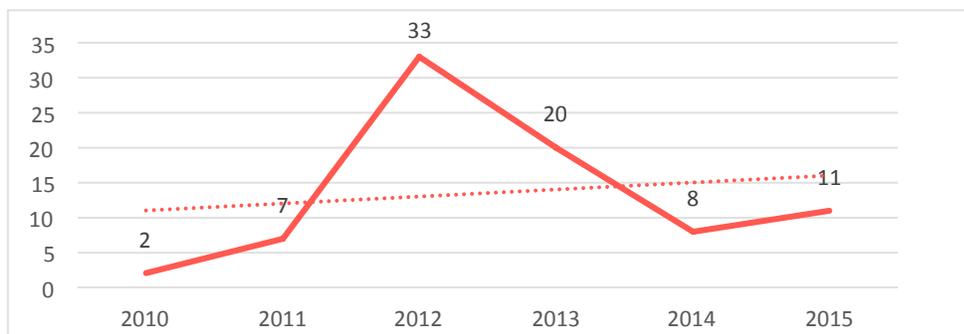

*Figure 4: Year of Publications*

Back again to Figure 4 accordingly, we can notice that 2012 has witnessed a big increased of number of publications ranged from 7 publications in 2011 to 33 publications by the end of the year 2012 which highlights the big interest of healthcare analytics topics in 2012, also 2013, but in 2014 and 2015 these publications were decreased as we believe that the reason behind that is relies on the rapid growth of number of population and technology in the last 2 years, therefore healthcare matter has become more complicated and quite uninteresting topic which need for a huge projects and more money to be implemented and tested. However, 2015 is a year of many challenges and extraordinary year of healthcare industry, as it has many ups and downs as well as a lot of issues have raised in this year related to the healthcare analytics, especially for the health IT vendors and government.

### 4.5 Research Approaches:

Out of 81 studies included in this systematic review, 50 studies have adopted qualitative research approach, 19 studies used quantitative methods, and finally 12 studies used a mixture between qualitative and quantitative research approaches.

See Figure 5, it shows that papers in healthcare analytics area adopted qualitative research approach focused by discussing professionals and doctor's adoption for the healthcare systems, as well as understanding individual's behavior, skills, medical knowledge and relationship between patients and doctors in order to improve healthcare analytics results. we do strongly agree that qualitative study has really big effect in studying healthcare analytics, however we believe that also in addition to patient's knowledge and professionals healthcare technology systems adoption, but also quantitative studies are really important and have a big role in terms of understanding and utilizing theories and mathematical equations, which are highly required in order to achieve an accurate and correct health results to improve healthcare system in general and to add benefits into the predictions systems as specific.

### 4.6  Online Databases

This literature review has found a total number of 81 papers were considered highly relevant to the healthcare analytics after studying of seven databases, whereas 25 articles were repeated in different databases (see Figure 6).

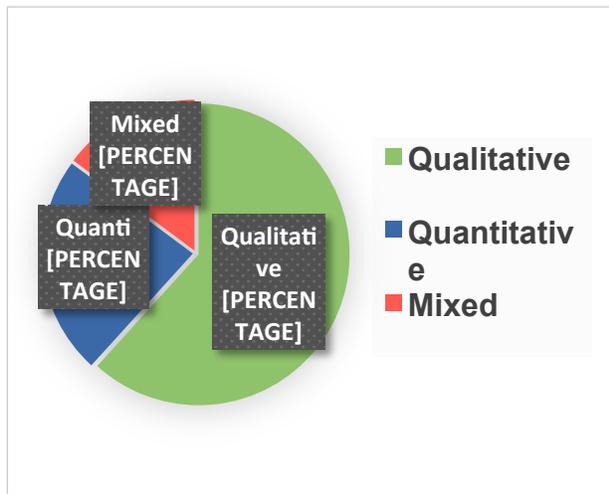
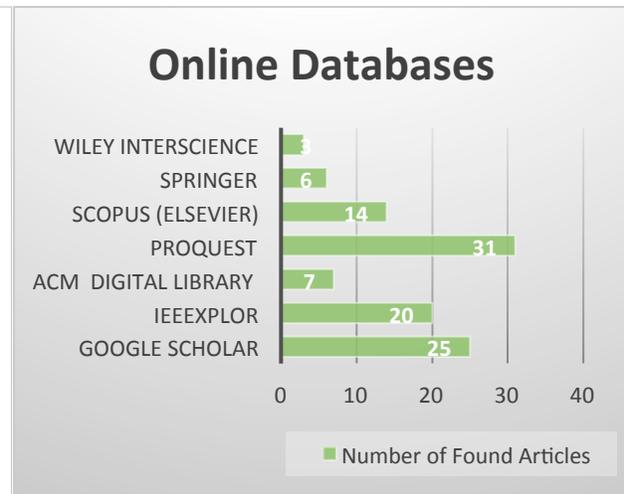

*Figure 5: Research Approaches*    *Figure 6: Online Databases*

As it shown in Table2 Google Scholar and ProQuest are the most popular databases, but if we compare between these two databases we can see that Google Scholar has listed a double number of papers than the papers listed in ProQuest, since 37,000 papers were listed in Google Scholar, while ProQuest listed 16,139 papers, although in my research we have found that ProQuest has proved to be a better database, since we have found 27 relevant papers to the topic of healthcare data analytics after filtering by text, while only 16 papers were founded in Google Scholar. This can be due to the inclusion of relevant papers in google Scholar but in multiple language options, which effects results after filtering by eliminating these papers, however ProQuest is only available in the English language. Also, ProQuest database provides full text access to articles, journals, eBooks and newspaper, but in Google Scholar some publications are only limited for a valued members or unless purchased.

We have found that IEEE and Springer databases had more relevant papers than ACM Digital Library and Scopus (Elsevier), for the reason that IEEE and Springer have included many number of academic papers than ACM Digital Library and Scopus. Finally, Wiley InterScience database is the least used database as a small number of papers came up with very few relevant articles after filtering by text. In conclusion, ProQuest and Google Scholar are the highest rated databases for this topic.

### 4.7  Geographic distribution:

Most of healthcare data analytics has been conducted in the United States and Europe, however there were some studies in Canada and very little in Asia. See Figure 7.

Figure 8 shows the distribution of research approaches in different areas. As we can see that United States and Canada have got kind of attention for both quantitative and qualitative studies, however in Europe qualitative studies have more attention than quantitative studies, as we believe this could be because quantitative studies would cost more than qualitative which is exceeded government budget, especially when the project fails.  Finally, Asia has been more into quantitative studies, and that could be because of huge number of populations in Asia which makes qualitative study a bit tough than quantitative, as well as qualitative studies publications written in English language which is not a common language in Asia. This study suggests if Europe and Asia can work together in order to improve qualitative studies in Asia, as well as supporting quantitative study needs in Europe.

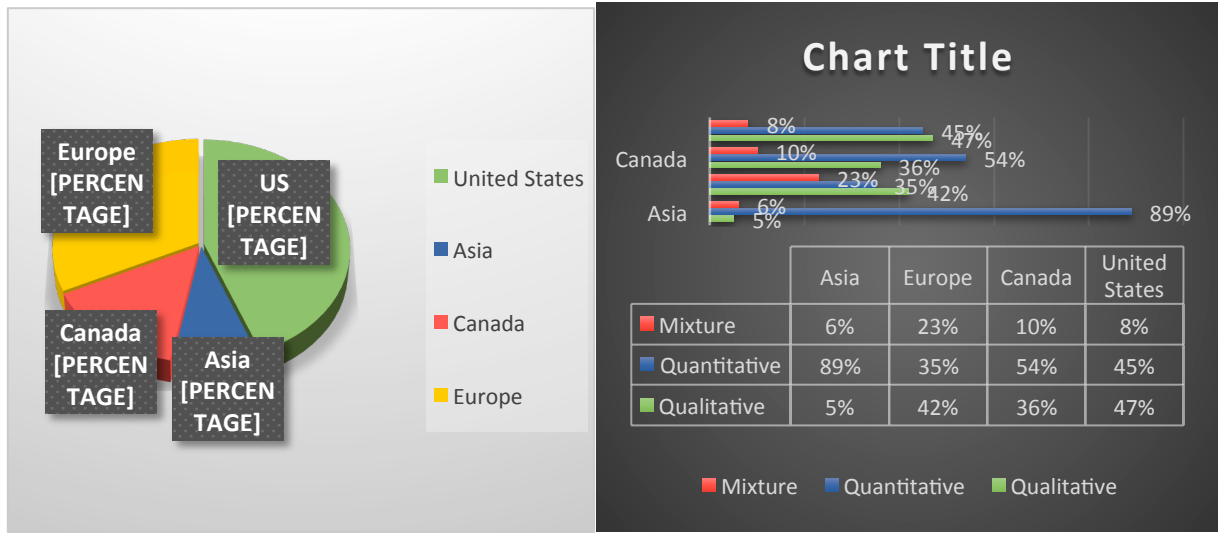

*Figure 7: Geographical Distribution of Studies in*　　*Figure 8: Distribution of Research Approaches geographical areas*

## 5　CONCLUSION & FUTURE WORK

In this paper we have listed 16 tools that are being useful in healthcare analytics. However, a research in this area is kind of difficult, as it's hard to push healthcare sectors and public to adopt a new data analytics techniques and tools, however we believe that highlighting some of the main factors is useful as it would help and provide a guidance with respect for healthcare data mining and analytics, as it would add a benefit to the healthcare decision systems and improve healthcare performance in the future, as well as pointing to some of the possible gaps in this topic. This work attends to give a good start for further studies in healthcare sectors as it demonstrates the positive impact of emerging between Information Technology field and Healthcare sectors. The next paragraphs will discuss the results found in this paper.

The publications seem to relate more to healthcare and information systems more than computer science, since computer science involves professionals and doctors only in healthcare analytics, paying less attention to involve individuals and patients, as we believe involving individuals will lead for better decision making and more accurate results for the future.

Most healthcare analytics papers were published in the US as it shown in geographical distribution in Figure 7, and linking this results to the research approach distribution in Figure 8, we can notice that most publications in healthcare analytics were quantitative studies, so if we take a look again, we can conclude and relate that to the reason of why papers were decreased in the last 2 years, because quantitative studies are very expensive and costly, as well as it is focusing on involving professionals in healthcare analytics systems rather than patients, which will effect on having many inaccurate studies results, and create many struggles with healthcare analytics and kills it's future, therefore researchers will stop publishing papers as they lost their spirit in healthcare analytics, so we recommend to involve individuals, as it would help society to adopt and improve healthcare data analytics systems, as well as running these systems efficiently and smoothly.

Also, from this study we have concluded that if Europe and Asia can collaborate and work together, then they can both get benefited from each other, in order to serve healthcare analytics researches in specific, and scientific academics studies in general, because English language in Asia is not popular, so English language in Asia and qualitative techniques can be improved more and more. Moreover, in this study Figure 3 has showed that healthcare information and diagnoses sharing is not an interesting area to be researched extensively, however we see that this area would help to increase patient's awareness and level of health education, so we suggest to have an equal attention to this area, as it helps patients to be more connected to the healthcare analytics systems in order to prevent and predict diseases and to keep them highly attached to the healthcare decision making systems.

Finally, this paper is proposing a technique that will promise to leverage large amount of healthcare data properly, since doctors and nurses will be able to determine diseases and risks easily like some certain types of cancer, diabetes and blood pressure, as well as provide needed treatment in the right time. Also, enhance doctor's decision making process by defining better care, developing drugs and vaccines along with a better treatment plan in order to reach patient satisfaction. Moreover, proposed technique will add a benefit of identifying risks early and mitigate it as much as possible. However, this study will need to push both doctors and patients to adopt new technique and collaborate together to reach high level of connection between both medical staff and patients in order to keep the system up to date and gather high quality of data. Also, this study will need from the individual to develop them selves and keep tracking their health conditions, but the problem is how to handle this with older people who are less attached and hard to convince to adopt new healthcare technologies and tools, as they consider this as a medical care issue involving medical staff and excluding their role in the medical care process.

Generally, in healthcare sectors data analytics is very important and essential topic, since all the previous benefits we mentioned could lead for better choice of medical care practice and prevent illnesses.

# 6   References

Abbott, PA & Coenen, A 2008, 'Globalization and advances in information and communication technologies: The impact on nursing and health', Nursing Outlook, vol. 56, no. 5, pp 238- 246.

Al-Azzawi, H. 2014. "Caradigm healthcare analytics." http://www.caradigm.com/media/68911/Caradigm-WP-Healthcare-Analytics-Jan-2014-US-EN.PDF Retrieved 09 August, 2015.

Bakshi, K. 2014, 'Considerations for big data: architecture and approaches', Aerospace conference, IEEE, pp. 1-7.

Bertsimas, D; Bjarnadóttir, M; Kryder, J; Pandey, R; Vempala, S; Wang, G. (2008), 'Algorithmic prediction of health-care costs', Operations Research, vol. 56, no. 6, pp. 1382-1557.

Beyer, K., Ercegovac, V., Gemulla, R., Balmin, A., Eltabakh, M., Ozcan, F. and Shekita, E. 2011. "Jaql: A Scripting Language for Large Scale Semi-Structured Data Analysis." http://web.cs.wpi.edu/~meltabakh/Publications/Jaql-PVLDB2011.pdf Retrieved 09 August, 2015.

Bhattacherjee, A & Hikmet, N 2007, 'Physicians' resistance toward healthcare information technology: a theoretical model and empirical test', European Journal of Information Systems: Including a Special Section on Healthcare Information, vol. 16, pp. 725–37.

Bradley, P., & Kaplan, J. (2010), 'Turning hospital data into dollars', HFM (Healthcare Financial Management), vol. 64, no. 2, pp. 64-68.

Brownstein, CA. & Wicks, P. 2010, 'The potential research impact of patient reported outcomes on osteogenesis imperfecta', Clin orthop relat Res, vol. 468, no. 10, pp. 5-2581.

Cannon, M., & Tanner, M. (2007), Healthy competition: What's holding back healthcare and how to free it, Cato Institute, Washington, D.C.

Castro, D 2007, Improving Health Care: Why a Dose of IT May Be Just What the Doctor Ordered, The information technology and innovation foundation, U.S.A.

Capriolo, E., Wampler, D. and Rutberglen, J. 2012. "Programming hive." http://www.reedbushey.com/99Programming%20Hive.pdf Retrieved 09 August, 2015.

Chordas, L. (2001), 'Building a better warehouse', Best's Review, vol. 101, no. 11, p. 117.

Confluent. 2015. "Avro." http://docs.confluent.io/1.0/avro.html Retrieved 09 August, 2015.

Conley, E., Owens, D., Luzio, S., Subramanian, M., Ali, A., Hardisty, A., & Rana, O. (2008), 'Simultaneous trend analysis for evaluating outcomes in patient-centred health monitoring services', Health Care Management Science, vol. 11, no. 2, pp. 152-66.

Datastax. 2013. "Comparing the hadoop distributed file system (HDFS) with the cassandra file system (CFS)." http://www.datastax.com/wp-content/uploads/2012/09/WP-DataStax-HDFSvsCFS.pdf Retrieved 09 August, 2015.

Dean, J. and Ghemawat, S., "MapReduce: simplified data processing on large clusters", Communication of The ACM - 50th Anniversary Issue: 1958 – 2008, 51, 1, January, 2008, pp 107-113.

Dieste, O., Grimán, A. and Juristo, N., "Developing search strategies for detecting relevant experiments", Empirical Software Engineering, 14, 5, October, 2009, pp 513-539.

Gabe, T. 2015. "Poverty in the United States: 2013." https://www.fas.org/sgp/crs/misc/RL33069.pdf Retrieved 09 August, 2015.

Ganeshkumar, P., Arun Kumar, S. and Rajoura, OP., "Evaluation of computer usage in healthcare among private practitioners of NCT Delhi", Stud Health Technol Inform, 169, 4, 2006, pp 960-964.

Good, Nathan A. 2013. "Big data serialization using Apache Avro with Hadoop Share serialized data among applications." http://www.ibm.com/developerworks/library/bd-avrohadoop/bd-avrohadoop-pdf.pdf Retrieved 09 August, 2015.

Green, S., "Systematic reviews and meta-analysis", Evidence-Based Medicine and Healthcare, 46, 3, 2005, pp 270-274.

Hanoch, Y., Miron-Shatz, T. & Himmelstein, M. 2012, 'Genetic testing and risk interpretation: how do women understand lifetime risk results?', Judgment and Decision Making, vol. 5, no. 2, pp. 23-116.

Herodotou, H., Lim, H., Luo, G., Borisov, N., Dong, L., Cetin, F.B. & Babu, S. 2011, 'A self-tuning system for big data analytics', Innovation data system research, Starfish, pp. 261-272.

Hortonworks. 2015. "Apache mahout." http://hortonworks.com/hadoop/mahout/ Retrieved 09 August, 2015.

HP. 2013. "Patient analytics." http://www.vertica.com/wp-content/uploads/2013/08/4aa4-7858ENW.pdf Retrieved 09 August, 2015.

IBM. 2011. "Cloud computing: building a new foundation for healthcare." https://www-05.ibm.com/de/healthcare/literature/cloud-new-foundation-for-hv.pdf Retrieved 09 August, 2015.

Imamura, T., Matsumoto, S., Kanagawa, Y., Tajima, B., Matsuya, S., Furue, M., & Oyama, H. (2007), 'A technique for identifying three diagnostic findings using association analysis', Medical & Biological Engineering & Computing, vol. 45, no. 1, pp. 51-59.

Jacob, S. 2012,'Young parkland physician makes a splash with predictive modeling software', D healthcare daily, Dallas, http://healthcare.dmagazine.com/2012/12/10/young-parkland-physician-makes-a-splash-with-predictive-modeling-software/ Retrieved 09 August, 2015.

Kim, J. 2013, 'The intersection of the quantified self-movement and big data', TechTarget, http://searchhealthit.techtarget.com/opinion/The-intersection-of-the-quantified-self-movement-and-big-data Retrieved 09 August, 2015.

Lakshman, A. and Malik, P., "Cassandra: a decentralized structured storage system", Operating Systems Review, 44, 2, April, 2010, pp 35-40.

Ledbetter, Craig S. and Morgan, Matthew W., "Toward best practice: leveraging the electronic patient record as a clinical data warehouse", Journal of Healthcare Information Management, 15, 2, Summer, 2001, pp 119-131.

LeRouge, C., Mantzana, V. & Vance Wilson, E. 2007, 'Healthcare information systems research, revelations and visions', European Journal of Information Systems, vol. 16, pp. 669 – 671.

LexisNexis, 'Healthcare clinical analytics', United States, viewed 4 April 2015,

Loginov, MV., Marlow, E., & Potruch, V. 2012, 'Predictive modeling in healthcare costs using regression techniques'. ARCH 2013.1 proceedings, pp. 1-32.

Luciano, SJ., Cumming, GP., Wilkinson, MD. & Kahana, E. 2013, 'The emergent discipline of health web science', J med internet res, vol. 15, no. 8, p. 166.

McAfee, A. & Brynjolfsson, E. 2012, 'Big data: the management revolution', Harvard business review, 1st October, https://hbr.org/2012/10/big-data-the-management-revolution/ar Retrieved 09 August, 2015.

McHorney, CA. (2009), 'The Adherence Estimator: a brief, proximal screener for patient propensity to adhereto prescription medications for chronic disease', Curr Med Res Opin, vol. 25, no, 1, pp. 215-38.

Miron-Shatz, T., Bowen, B., Diefenbach, M., Goldacre, B., Muhlhauser, I. & Smith, RSW. 2011, 'Barriers to health information and building solutions', Better decisions: envisioning health care, vol. 6, no. 1, pp. 191-212.

Murphy, K. 2013, 'What about healthcare must change for analytics to flourish?', Ehrintelligence, 29 August, https://ehrintelligence.com/2013/08/29/what-about-healthcare-must-change-for-analytics-to-flourish/ Retrieved 09 August, 2015.

Nori, S. 2014. "5 ways hadoop can help healthcare organizations and you" http://www.smartdatacollective.com/sameernori/282021/5-ways-hadoop-can-help-healthcare-organizations-and-you Retrieved 09 August, 2015.

Obenshain, M. (2004), 'Application of data mining techniques to healthcare data', Infection Control and Hospital Epidemiology, vol. 25, no. 8, pp. 690-695.

Peši , S., Stankovi , T. and Jankovi , D., "Benefits of using OLAP versus RDBMS for data analyses in health care information systems", Electronics, 13, 2, December, 2009, pp 56-60.

Philips. 2015. "Open a world of cloud-based collaborative care." http://www.usa.philips.com/healthcare-innovation/about-health-suite Retrieved 09 August, 2015.

Plattner, H. & Zeier, A. 2011, In-memory data management: an inflection point for enterprise applications, Springer, Heidelberg, Germany.

Powell, James E. 2014. "Q&A: Advanced data visualization: from atomic data to big data." https://tdwi.org/articles/2014/09/02/advanced-data-visualization.aspx Retrieved 09 August, 2015.

Raden ,N. 2010. "Get analytics right from the start." file:///C:/Users/Mohammad%20al%20khatib/Downloads/X/Analytics-from-the-Start-WP.pdf Retrieved 09 August, 2015.

Raja, U., Mitchell, T., Day, T. and Michael Hardin, J. 2014. "Text mining in healthcare. Applications and opportunities." http://www.researchgate.net/publication/24182770_Text_mining_in_healthcare._Applications_and_opportunities Retrieved 09 August, 2015.

Ried K, "Interpreting and understanding meta-analysis graphs--a practical guide", Aust Fam Physician, 35, 8, August, 2006, pp 635-638.

Russom, P. 2011, 'Executive summary big data analytics', Tdwi best practice report, Renton, Washington.

Shmueli, G., & Koppius, O. (2011), 'Predictive analytics in information systems research', MIS Quarterly, vol. 35, no. 3, pp. 553-572.

Shvachko, K., Kuang, H., Radia, S, and Chansler, R. 2010. "The hadoop distributed file system" http://zoo.cs.yale.edu/classes/cs422/2014fa/readings/papers/shvachko10hdfs.pdf Retrieved 09 August, 2015

Siemens, G. 2004, 'Connectivism: A learning theory for the digital age', http://www.elearnspace.org/Articles/connectivism.htm Retrieved 09 August, 2015.

Swan, M. 2012, 'Crowdsourced health research studies: an important emerging complement to clinical trials in the public health research ecosystem'. J med internet res, vol. 14, no. 2, p. 46.

Taylor, J. 2010. 'Transforming healthcare delivery with analytics: Improving outcomes and point-of-care decisioning', Decision management solutions, ftp://public.dhe.ibm.com/software/data/sw-library/infosphere/analyst-reports/Transforming_Healthcare_Delivery.pdf Retrieved 09 August, 2015.

Turner, MR., Wicks, P., Brownstein, CA., Massagli, MP., Toronjo, M. & Talbot K. 2011, 'Concordance between site of onset and limb dominance in amyotrophic lateral sclerosis', J neurol neurosurg psychiatry, vol. 82, no. 8, pp. 4-853.

Vertica. 2010. "The vertica analytic database technical overview white paper." http://www.vertica.com/wp-content/uploads/2011/01/VerticaArchitectureWhitePaper.pdf Retrieved 09 August, 2015.

Wager, K.A., Lee, F.W. & Glaser, J.P. 2005, ManagingHealth Care Information Systems: A Practical Approach for Health Care Executives, Wiley, United states of America.

Walker, M., Fonda, SJ., Salkind, S. & Vigersky, R. 2012, 'Advantages and disadvantages of real-time continues glucose monitoring in people with type 2 diabetes', Blood glucose monitoring, vol. 8, no. 1, p. 25.

Webster, C. 2011. "Clinical intelligence, complex event processing and process mining in process-aware EMR / EHR BPM systems." http://chuckwebster.com/2011/07/clinical-intelligence/clinical-intelligence-complex-event-processing-process-mining-process-aware-emr-ehr-bpm-systems Retrieved 09 August, 2015.

Wongsuphasawat, K., Gomez, J. A. G., Plaisant, C., Wang, T. D., Shneiderman, B., and Taieb-Maimon, M., "LifeFlow: visualizing an overview of event sequences", Proc. ACM SIGCHI Conference, ACM Press, New York, May 2011, pp 1747-1756.

Wulff, F. 2013. "Presto." https://prestodb.io/ Retrieved 09 August, 2015.